\documentclass[letterpaper]{article} % DO NOT CHANGE THIS

\ifdefined\aaaianonymous
    \usepackage[submission]{aaai2026}  % Anonymous submission version
\else
    \usepackage{aaai2026}              % Camera-ready version
\fi

\usepackage{times}  % DO NOT CHANGE THIS
\usepackage{helvet}  % DO NOT CHANGE THIS
\usepackage{courier}  % DO NOT CHANGE THIS
\usepackage[hyphens]{url}  % DO NOT CHANGE THIS
\usepackage{graphicx} % DO NOT CHANGE THIS
\usepackage{natbib}  % DO NOT CHANGE THIS AND DO NOT ADD ANY OPTIONS TO IT
\usepackage{caption} % DO NOT CHANGE THIS AND DO NOT ADD ANY OPTIONS TO IT
\usepackage{algorithm}
\usepackage{algorithmic}
\usepackage{amsmath}
\usepackage{amssymb}
\usepackage{booktabs}
\usepackage{tikz}
\usetikzlibrary{arrows.meta,positioning,shapes.geometric,calc}

\usepackage{newfloat}
\usepackage{listings}
\DeclareCaptionStyle{ruled}{labelfont=normalfont,labelsep=colon,strut=off}
\floatstyle{ruled}
\newfloat{listing}{tb}{lst}{}
\floatname{listing}{Listing}

\title{TAG: A Lightweight Framework for\\Test-Driven Agentic Artifact Generation}

\author{
    Yaniv Melamed,\textsuperscript{\rm 1}
    Yoni Zukerman,\textsuperscript{\rm 1}
    Michal Shechter,\textsuperscript{\rm 1}\\
    Miri Weissler,\textsuperscript{\rm 1}
    Ashwin Patil,\textsuperscript{\rm 1}
    Hani Neuvirth-Telem\textsuperscript{\rm 1}
}
\affiliations{
    \textsuperscript{\rm 1}Microsoft
}

\nocopyright

\begin{document}
\maketitle
\begingroup
\renewcommand{\thefootnote}{}
\footnotetext{\raggedright Correspondence: \texttt{yanivmelamed@microsoft.com}.}
\endgroup

\begin{abstract}
Generating structured artifacts with Large Language Models - e.g.\ database queries, threat framework mappings, entity schemas - is relatively straightforward; however, making them reliable enough for production deployments presents challenges.
We present TAG, a lightweight framework based on a core principle: \textit{LLMs generate, we validate}.
This reframing shifts responsibility from generation quality to validation rigor.
The framework rests on three key attributes:
First, \textbf{test driven generation}: when tests fail, the LLM receives indicative error messages that expose why the output failed, enabling the LLM to understand its mistakes and refine subsequent attempts.
Second, \textbf{deterministic and LLM-based tests}: deterministic tests catch heuristics that can be programmatically verified (schema, syntax, cross-reference), while LLM-based tests evaluate nuanced semantic and delicate features that resist programmatic inspection (intent alignment, logical consistency, domain correctness).
Third, \textbf{expert-distilled judges}: LLM-based tests are calibrated to distill and replicate human expert decision distribution, transforming manual human quality gates into scalable, reusable evaluation proxies that reflect professional-grade validation standards.
We demonstrate the framework on three artifact types in the security domain - KQL query generation, MITRE ATT\&CK mapping, and entity mapping - deployed in production at Microsoft Sentinel.
We believe this framework can be applied beyond security to other artifact generation tasks, providing a path to reliable, high-quality outputs without sacrificing the efficiency gains of LLM generation.
\end{abstract}

% ============================================================
\section{Introduction}
\label{sec:introduction}
% ============================================================

Across many professional domains, experts spend significant time crafting structured artifacts that run in production systems and serve customers at scale.
A threat analyst authors detection rules that continuously monitor for advanced persistent threats across an enterprise with millions of endpoints.
A compliance engineer maps regulatory requirements to executable policy rules that enforce data residency and encryption standards at scale.
A data engineer designs ETL pipelines that transform raw telemetry into normalized schemas powering downstream ML models and business intelligence.
An infrastructure engineer provisions cloud environments using infrastructure-as-code templates that reliably deploy multi-region systems to production.
A healthcare informaticist maps clinical observations to standardized medical coding frameworks that ensure interoperability across hospital systems.
A security architect encodes threat modeling results into entity schemas that automatically classify assets and risks in security platforms.

These artifacts share a common profile: they are \emph{authored once} by a domain expert, \emph{validated carefully} against correctness criteria, and then \emph{deployed broadly} to production where they operate autonomously.
The authoring process is slow and expensive - it requires deep domain knowledge, iterative refinement, and manual quality assurance - but the deployment is high-leverage: a single well-crafted artifact can serve an entire customer base.

Large Language Models have made it remarkably easy to \emph{generate} such artifacts.
Given a natural-language specification, an LLM can produce a database query, a configuration file, a mapping to a standardized framework, or an entity schema in seconds.
Making those artifacts \emph{reliable enough to deploy}, however, remains hard.
A single hallucinated column name, an over-broad filter, or a semantically incorrect mapping is enough to render a generated artifact useless - or worse, silently wrong in production.
The dominant response has been to add human review: experts inspect each generated artifact before deployment.
But this reintroduces the bottleneck that automation was meant to eliminate.

We observe that the quality problem is not primarily a \emph{generation} problem - modern LLMs produce reasonable first drafts.
It is a \emph{verification} problem: knowing whether the output is correct, and what to do when it is not.
This reframing led us to a test-driven approach in which the test suite, not the model, defines the acceptance criteria.

We present TAG, a lightweight framework for artifact generation built around three ideas:

\begin{enumerate}
\item \textbf{Test-driven agentic generation.}
An autonomous agent equipped with tools and tasked with producing structured artifacts. After autonomous tool usage for exploration and information gathering, the agent produces a final artifact. That artifact is then subjected to an explicit test suite that encodes correctness criteria.
Failures in one of the tests are fed back to the agent with indicative error messages, creating an iterative refinement loop.

\item \textbf{Programmatic and semantic testing.}
Artifact correctness has two facets.
Some properties - e.g. schema conformance, syntax validity  - are rigorously definable and checked programmatically.
Others - e.g. does this query capture the intended logic? is this mapping conceptually appropriate? - require expert-like judgment.
Our framework addresses both: deterministic tests encode precise constraints, while LLM-based tests handle semantic evaluations that resist formal specification.

\item \textbf{Expert-calibrated LLM judges.}
LLM-based tests are only as good as their alignment with human expert judgment.
We introduce a calibration mechanism that optimizes judge prompts against a dataset of expert-labeled examples, tuning the prompt to replicate the expert's decision profile - prioritizing low false-positive rates to ensure that bad artifacts are reliably rejected.
\end{enumerate}

While the framework is domain-agnostic by design, we demonstrate and evaluate it in the security domain, where we apply it to three artifact types: KQL query generation, MITRE ATT\&CK mapping, and entity mapping.
The system is deployed in production at Microsoft Sentinel, producing hundreds of validated behavior rules across multiple data sources.

% ============================================================
\section{Test-Driven Agentic Generation}
\label{sec:framework}
% ============================================================

The premise is straightforward: give the model autonomy to explore data and produce artifacts, but make acceptance conditional on passing an explicit test suite.
When tests fail, indicative error messages are fed back as part of the conversation, creating a self-correcting loop.
The agent generates; the tests gate.

\subsection{The Agent Loop}
\label{sec:agent-loop}

\begin{figure}[t]
\centering
\includegraphics[width=\columnwidth]{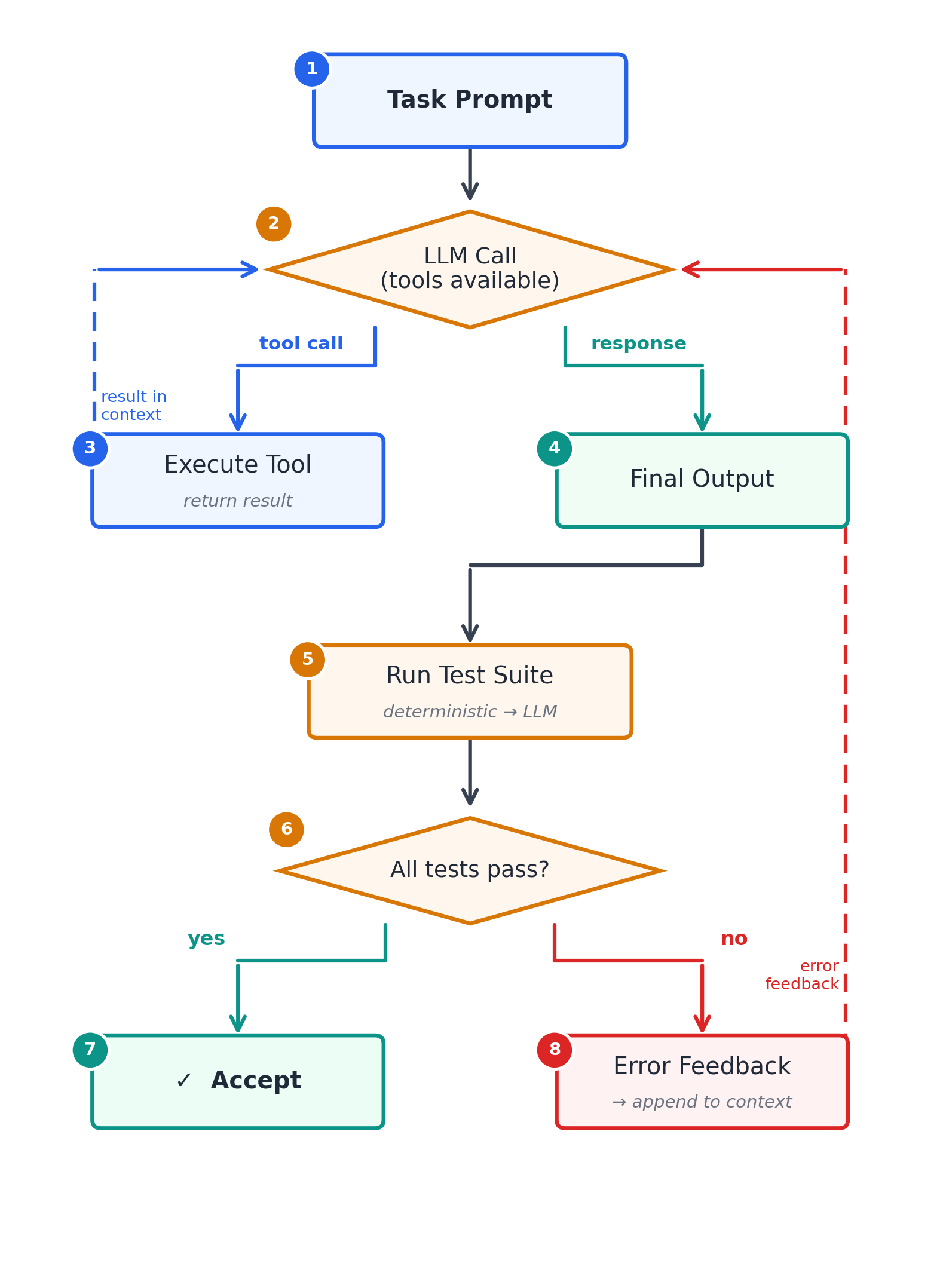}
\caption{Test-driven agentic generation loop. The agent alternates between tool-call exploration (left loop) and artifact generation. Failed tests produce actionable error messages appended to the conversation (right loop), driving iterative self-correction.}
\label{fig:agent-flow}
\end{figure}

An agent instance is defined by system prompt instructions, a task prompt, a set of tools for exploration, and a test suite encoding acceptance criteria.
The core loop (Algorithm~\ref{alg:tdg-loop}) starts with the initial prompt. The agent autonomously uses tools to explore and gather information, then produces a candidate artifact. We emphasize that the decision of which tools to call, when to call them, and when to stop exploring and generate is entirely up to the model. After the agent generates an artifact, the framework runs the test suite. If all tests pass, the artifact is accepted and registered and the loop terminates successfully. If any test fails, the error message is appended to the conversation context as a user-role message, and the loop continues - allowing the agent to understand its mistake, rerun some tool calls, and refine its approach in subsequent iterations. Figure~\ref{fig:agent-flow} illustrates the complete flow.

\begin{algorithm}[t]
\caption{Test-Driven Agentic Generation Loop}\label{alg:tdg-loop}
\begin{algorithmic}[1]
\REQUIRE Task prompt $p$, tools $T$, test suite $S$, max calls $N$, max retries $R$
\ENSURE Validated artifact $a$ or $\bot$
\STATE $ctx \gets [p]$, $calls \gets 0$, $fails \gets 0$
\WHILE{$calls < N$}
    \STATE $resp \gets \textsc{LLM}(ctx, T)$
    \IF{$resp$ contains tool calls}
        \STATE $results \gets \textsc{ExecuteTools}(resp.tools)$
        \STATE $ctx \gets ctx \cup resp \cup results$
    \ELSE
        \STATE $err \gets \textsc{RunTests}(resp.output, S)$
        \IF{$err = \emptyset$}
            \RETURN $resp.output$
        \ELSE
            \STATE $ctx \gets ctx \cup resp \cup err$
            \STATE $fails \gets fails + 1$
            \IF{$fails \geq R$}
                \RETURN $\bot$
            \ENDIF
        \ENDIF
    \ENDIF
    \STATE $calls \gets calls + 1$
\ENDWHILE
\STATE \RETURN $\bot$ \COMMENT{Max calls exceeded}
\end{algorithmic}
\end{algorithm}

Three conditions terminate the loop:
\begin{itemize}
\item \textbf{Success}: all tests pass and the output parses into the target schema.
\item \textbf{Max calls}: after $N$ LLM calls (we use $N{=}60$) without success, the run terminates.
\item \textbf{Max test failures}: after $R$ consecutive failures (typically $R{=}10$), the run terminates.
\end{itemize}

Additionally, the model has access to a \texttt{give\_up} tool: if it determines that the task is fundamentally unachievable (e.g., required data does not exist), it can explicitly abort with a reason.
This is a deliberate design choice - not every task is achievable, and recognizing this early saves compute.

%  -  Agent Flow Figure  - 

\subsection{Test Suite}
\label{sec:tests}

Tests are the mechanism through which domain knowledge enters the framework.
Each test receives the model's output and an optional supplementary data for evaluation. 
A test either returns \texttt{pass} or an \texttt{error message} that is fed back to the agent.

Tests fall into two categories, executed sequentially.
\textbf{Deterministic tests} are fast, free, and structural.
They validate JSON schema conformance, check that referenced identifiers exist in a source schema, enforce naming conventions, verify value ranges, and ensure uniqueness constraints.
These tests catch the majority of errors at negligible cost.

\textbf{LLM-based tests} are slower and semantic.
A separate LLM instance (acting as an evaluator, not the generator) assesses different semantic properties of the output. For example,
whether the artifact aligns with the original intent, whether a classification is conceptually correct, or whether the output contains problematic content.
Each evaluation returns a verdict with an explanation, which in case of test failure is fed back to the agent as an error message.
The \emph{Expert-Calibrated LLM Judges} section below describes how these evaluators are calibrated against human expert judgment.

Two design choices keep evaluation cost-efficient.
\emph{First-failure-stops}: tests run in a fixed order; the first failure aborts the suite and returns the error to the model.
A trivially malformed output never triggers an expensive LLM evaluation.
\emph{Actionable error messages}: each test produces feedback that tells the model exactly what went wrong and which dimension was evaluated, enabling targeted correction rather than undirected retries.

% ============================================================
\section{Expert-Calibrated LLM Judges}
\label{sec:judges}
% ============================================================

\subsection{Why Calibration Is Needed}

Some artifact properties e.g., intent alignment, conceptual correctness, usefulness, cannot be checked programmatically; they require the kind of judgment a human expert would apply.
In a manual workflow, a domain expert reviews each artifact before it ships.
This is the bottleneck we set out to eliminate: expert review does not scale.

LLM-based tests address this by delegating semantic evaluation to a separate model instance acting as a judge.
But this raises a fundamental question: \emph{how do we know the judge is reliable?}
If the judge accepts artifacts that a human expert would reject, we have automated the bottleneck away while also removing the quality gate.
The goal of calibration is to ensure the judge's decision distribution matches the human expert's - meaning that if a domain expert would reject an artifact on a given dimension, the judge should reject it too.
We introduce a simple pipeline for optimizing judge prompts against a dataset of expert-labeled examples, with the key insight that the optimization should explicitly preserve the expert's decision profile rather than merely maximizing accuracy.

\subsection{Calibration Method}
\label{sec:calibration-method}

The calibration process takes as input a set of expert-labeled examples and an initial judge prompt, and produces a refined prompt that replicates the expert's scoring behavior (Algorithm~\ref{alg:calibration}).
Critically, each evaluation dimension has its own judge prompt and is calibrated independently.
For example, an artifact may be evaluated on \emph{usefulness} (is this output practically valuable?) and \emph{correctness} (does the output faithfully reflect the specification?) -- these are separate judges with separate prompts, each calibrated against its own set of expert labels.
The full calibration pipeline runs once per dimension, yielding $N$ independently optimized judge prompts for $N$ evaluation dimensions.

\paragraph{Step 1: Expert labeling.}
A domain expert reviews a representative batch of generated artifacts.
For each artifact, the expert provides a binary verdict (valid/not valid) and a free-text explanation of their reasoning, independently for each evaluation dimension.
The same artifact may be labeled ``valid'' on one dimension and ``not valid'' on another.

\paragraph{Step 2: Dataset splitting.}
For each dimension, the labeled set is split into training (60\%), validation (20\%), and test (20\%) partitions using stratified sampling to preserve the expert's label distribution.

\paragraph{Step 3: Iterative prompt refinement.}
For each dimension independently, the optimization loop proceeds over multiple epochs with randomized minibatches from that dimension's training set.
For each minibatch:
\begin{enumerate}
\item The current judge prompt is applied to each example. To reduce variance, each judgment is made by \emph{majority vote} over $k$ independent LLM calls (we use $k{=}3$).
\item Predictions are compared to expert labels. Disagreements are annotated with the expert's explanation and categorized by error type.
\item If disagreements exist, a \emph{meta-optimizer} -- a separate, high-capability LLM -- receives the current prompt, the disagreement cases with full context, and instructions to revise the evaluation criteria while:
  (a)~prioritizing false-positive reduction,
  (b)~avoiding overfitting to the batch,
  (c)~preserving the overall decision distribution, and
  (d)~modifying only the evaluation criteria, not the output format.
\end{enumerate}

\paragraph{Step 4: Checkpoint selection.}
After each epoch the updated prompt is evaluated on that dimension's validation set.
The final prompt is selected as the checkpoint with the best validation accuracy subject to the lowest false-positive rate.

\paragraph{Step 5: Held-out evaluation.}
The selected prompt is evaluated on the held-out test set for an unbiased estimate of judge--expert agreement on that dimension.

\begin{algorithm}[t]
\caption{Judge Prompt Calibration (per dimension)}\label{alg:calibration}
\begin{algorithmic}[1]
\REQUIRE Expert labels $D$, initial prompt $J_0$, epochs $E$, votes $k$
\ENSURE Calibrated prompt $J^*$
\STATE Split $D \to D_{train}, D_{val}, D_{test}$ (stratified)
\STATE $J \gets J_0$
\FOR{$e = 1$ to $E$}
    \FOR{each minibatch $b \subset D_{train}$}
        \STATE $preds \gets \textsc{MajorityVote}(\textsc{Judge}(J, b),\; k)$
        \STATE $errors \gets \{(x, pred, label) : pred \neq label\}$
        \IF{$errors \neq \emptyset$}
            \STATE $J \gets \textsc{MetaOptimize}(J, errors)$
            \STATE \quad \COMMENT{Minimize FP; preserve distribution; avoid overfitting}
        \ENDIF
    \ENDFOR
    \STATE $acc, fpr \gets \textsc{Evaluate}(J, D_{val})$
    \STATE Save checkpoint $(J, acc, fpr)$
\ENDFOR
\STATE \RETURN $J^* = \arg\min_{J} fpr(J)$ among highest-accuracy checkpoints
\end{algorithmic}
\end{algorithm}

\subsection{Design Choices}
\label{sec:calibration-choices}

Our calibration approach is conceptually related to TextGrad~\cite{yuksekgonul2025textgrad}, which performs automatic prompt optimization using LLM-generated textual gradients.
We opted for a simpler, manual iterative refinement: the meta-optimizer proposes prompt revisions based on explicit disagreement cases, and these revisions are guided by human-specified priorities (minimize false positives, preserve distribution).
This reflects our setting where the number of evaluation dimensions is small (typically 2-3 per artifact type) and the cost of calibration is dominated by expert labeling, not by prompt search.

The deliberate prioritization of false-positive reduction over overall accuracy reflects an asymmetry in production costs: a bad artifact that passes the judge ships to all customers and may cause harm, while a good artifact that is rejected can be regenerated or reviewed manually.
We accept a higher false-negative rate as the price of conservative gatekeeping.

% ============================================================
\section{Multi-Stage Artifact Pipelines}
\label{sec:pipelines}
% ============================================================

When a production artifact requires multiple properties - each with its own quality criteria - the framework extends naturally to a multi-stage pipeline.
Each stage is an independent instance of the agent loop (Figure~\ref{fig:agent-flow}), tasked with generating or refining one property of a seed artifact.
Stages are chained sequentially: each receives the accumulated output of previous stages as part of its input context.

This architecture offers three advantages.
First, \emph{specialization}: each stage has its own tools, tests, and prompts, optimized for one generation task.
Second, \emph{fault isolation}: a failure at one stage does not invalidate earlier work; the artifact exits the pipeline at the failing stage.
Third, \emph{progressive filtering}: the pipeline acts as a quality funnel - only artifacts that survive every stage's test suite ship to production.

The practical consequence is that earlier stages, which address harder generation tasks, absorb the majority of attrition, while later stages operate on already-validated partial artifacts and achieve high pass rates.
The \emph{Application} section demonstrates this pattern on a production pipeline.

% ============================================================
\section{Application: Security Behavior Rules}
\label{sec:application}
% ============================================================

We demonstrate the framework on the generation of security behavior rules - structured detection artifacts deployed in production at Microsoft Sentinel across 9 heterogeneous data sources~\cite{shechter2026ueba}.

\subsection{What Is a Behavior Rule}

A \emph{behavior rule} is a detection artifact that continuously monitors security telemetry and surfaces instances of specific activity patterns.
Unlike an alert (which implies a verdict), a behavior is a neutral observation: ``this entity performed these actions in this time window.''

A complete behavior rule comprises four components, each a distinct structured artifact:
\begin{itemize}
\item A \textbf{KQL query} that runs against a log table (e.g., \texttt{AWSCloudTrail}, \texttt{GCPAuditLogs}) and outputs matching records.
\item A \textbf{MITRE ATT\&CK mapping}~\cite{mitre2024attack}: technique and tactic identifiers from the ATT\&CK framework.
\item An \textbf{entity mapping}: structured extraction of actors, targets, and resources from query output columns, with typed roles (e.g., \texttt{Account/Subject}, \texttt{IP/Source}).
\item A \textbf{dynamic description template}: natural-language text with placeholders filled per instance.
\end{itemize}

The system operates across multiple data sources - each with different schemas, column names, and event semantics.

\subsection{Pipeline Instantiation}

The behavior generation pipeline instantiates the \emph{Multi-Stage Artifact Pipelines} architecture with three core generation stages - KQL query, MITRE mapping, and entity mapping - followed by two lightweight downstream stages (dynamic description generation and volume verification).
Table~\ref{tab:tools} summarizes tool and test assignments for the three core stages; the downstream stages use minimal tooling and are not detailed here.

\paragraph{Stage 1: KQL query generation.}
given a natural-language hypothesis of a behavior rule (which generated in a pre processing step by a separate LLM agent), the first layer agent produces a KQL query that captures the specified behavior.
The agent inspects table schemas (\texttt{get\_table\_schema}), executes exploratory queries against live data (\texttt{run\_kusto\_query}), and profiles output columns (\texttt{columns\_report}).
The test suite includes 12 deterministic tests (e.g., schema validation, syntax, naming conventions, execution against live data, compatibility checks) and multiple LLM-based tests evaluating, for example, intent alignment and usefulness.
These judges are calibrated via the method described in \emph{Expert-Calibrated LLM Judges}.

\paragraph{Stage 2: MITRE ATT\&CK mapping.}
Rather than injecting the full ATT\&CK matrix into the prompt, the agent browses it incrementally: listing all techniques, drilling into sub-techniques, reading descriptions, and querying tactics.
Two deterministic tests validate schema and technique-ID existence; three LLM judges assess mapping correctness.

\paragraph{Stage 3: Entity mapping.}
The agent discovers entity types, schemas, and role definitions, then maps query output columns to typed entities.
One deterministic test validates the structure; three LLM judges evaluate semantic correctness and data safety.
The cross-source challenge is acute: an account entity identifier can be one column in one data source and a completely different column in another; the agent must explore and understand the schema to produce a correct mapping.

\begin{table}[t]
\caption{Tool and test assignments per pipeline stage. Built-in tools (notebook, recall, give\_up) are available to all agents.}
\label{tab:tools}
\centering
\small
\begin{tabular}{lp{4.0cm}cc}
\toprule
\textbf{Stage} & \textbf{Domain Tools} & \textbf{Det.} & \textbf{LLM} \\
\midrule
KQL Query      & \texttt{get\_table\_schema}, \texttt{run\_kusto\_query}, \texttt{columns\_report} & 12--14 & 2--3 \\
\midrule
MITRE Map.     & \texttt{get\_all\_techniques}, \texttt{get\_sub\_techniques}, \texttt{get\_tactics}, \texttt{get\_description} & 2 & 3 \\
\midrule
Entity Map.    & \texttt{get\_entities\_overview}, \texttt{get\_entity\_schema}, \texttt{get\_entity\_role\_defs} & 1 & 3 \\
\bottomrule
\end{tabular}
\end{table}

\subsection{End-to-End Production Funnel}
\label{sec:funnel}

In production, the pipeline operates on a \emph{hypothesis bank} - a collection of natural-language behavior hypotheses generated upstream by a separate LLM module.
Each hypothesis describes a candidate security behavior.
The three core stages described above are followed by the two lightweight downstream stages, forming an end-to-end refinement funnel.
Each stage acts as a quality gate: a hypothesis that fails to produce a valid artifact at any stage is discarded from further processing.

\begin{figure}[t]
\centering
\includegraphics[width=\columnwidth]{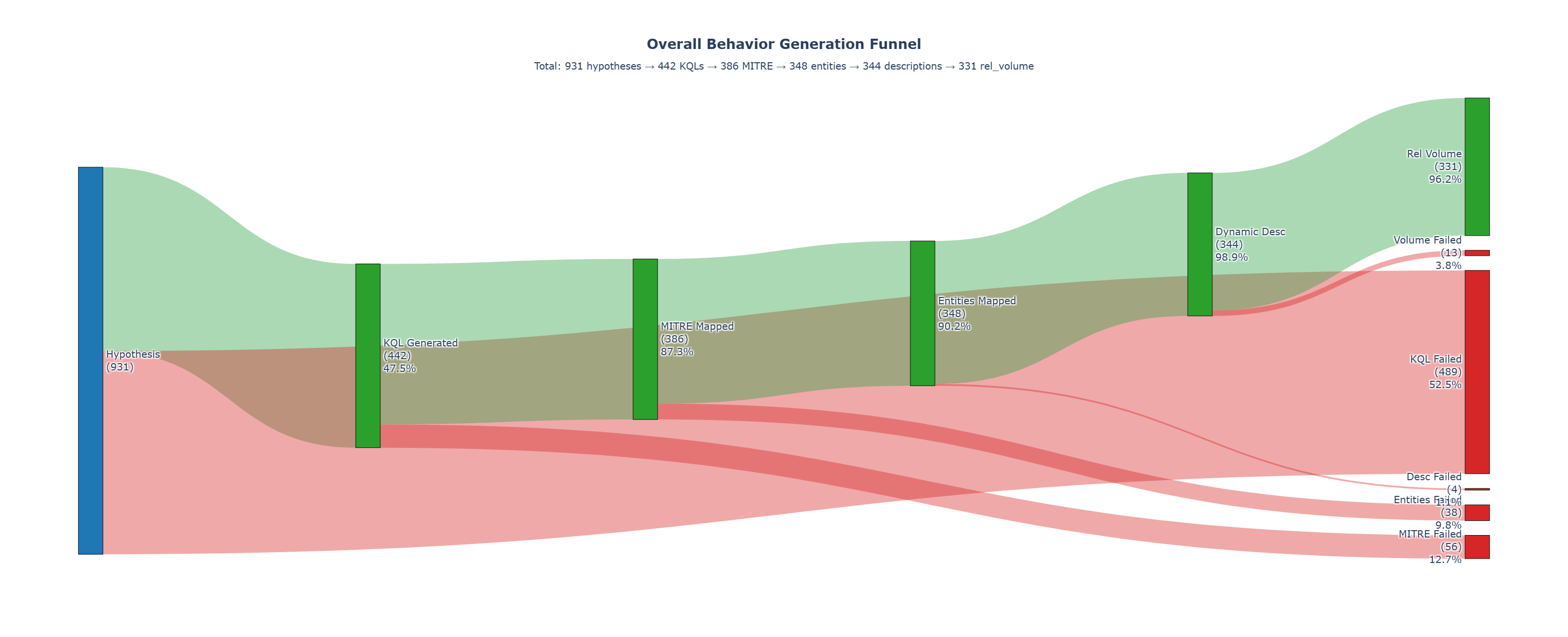}
\caption{Production funnel across 931 hypotheses. Green flows indicate hypotheses passing each stage; red flows indicate stage-specific failures. KQL generation is the primary bottleneck (52.5\% of hypotheses discarded); later stages achieve ${>}87\%$ pass rates.}
\label{fig:funnel}
\end{figure}

Figure~\ref{fig:funnel} shows the production funnel across 931 hypotheses.
KQL generation is the primary bottleneck, discarding 52.5\% of initial hypotheses - queries must execute against live telemetry and pass 14+ deterministic and semantic tests.
Subsequent stages exhibit progressively higher pass rates: 87.3\% for MITRE mapping, 90.2\% for entity mapping, 98.9\% for description generation, and 96.2\% for volume verification.
The end-to-end yield is 331 production-ready behavior rules (35.6\%).

This attrition pattern - steep early, shallow late - is a direct consequence of the fault-isolation property described in \emph{Multi-Stage Artifact Pipelines}: later stages operate on already-validated partial artifacts and face fewer failure modes.

\subsection{Ablation Study}
\label{sec:ablation}

To quantify each component's contribution, we conduct a one-component-at-a-time ablation on the MITRE ATT\&CK mapping stage across 50 behavior rules.
We select this stage because it exercises both deterministic and LLM-judge tests, uses domain-specific browsing tools, and generates a semantically rich artifact.
Four variants are evaluated: (1)~\emph{w/o Self-Correction} - single-shot generation with no retry loop; (2)~\emph{w/o LLM Judges} - only deterministic tests active during generation; (3)~\emph{w/o Domain Tools} - MITRE browsing APIs removed; and (4)~\emph{w/o Quality Gates} - all generation-time tests disabled.
To ensure a uniform quality baseline, all outputs are evaluated \emph{post-hoc} by running the full five-test suite (two deterministic, three LLM-judge) independently of whatever tests were active during generation.
Note that the full pipeline achieves 80\%, not 100\%, because the generation-time retry budget is bounded; some artifacts exhaust retries without satisfying every test.

\begin{figure}[t]
\centering
\includegraphics[width=\columnwidth]{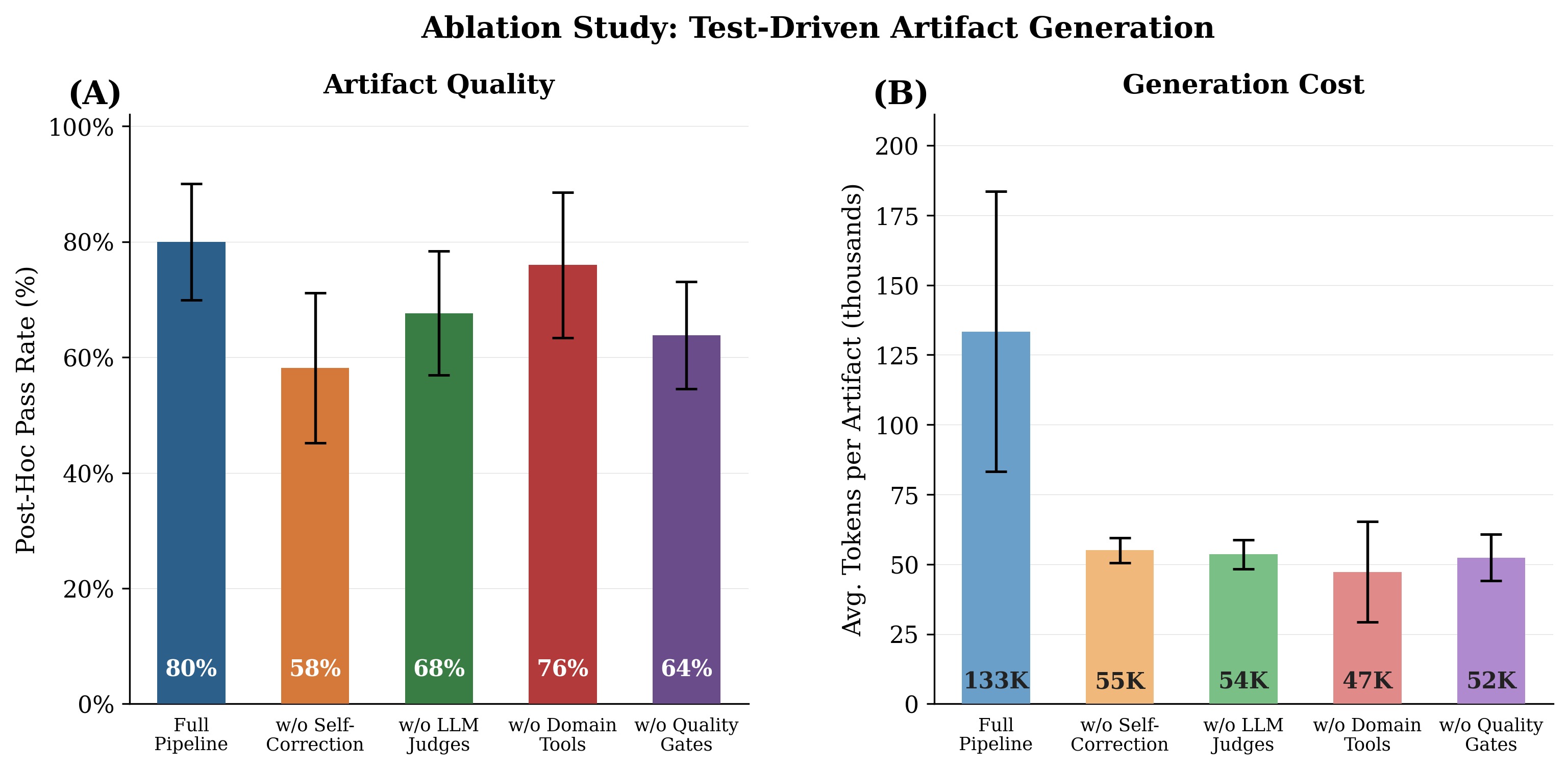}
\caption{Ablation on MITRE ATT\&CK mapping across 50 behaviors.
\textbf{(A)}~Post-hoc pass rate on the full five-test suite.
\textbf{(B)}~Mean token cost per artifact.
Removing self-correction causes the largest quality drop; the full pipeline costs ${\sim}2.5\times$ more.
Error bars: SD across four folds.}
\label{fig:ablation}
\end{figure}

Figure~\ref{fig:ablation} summarizes the results.
Self-correction is the most impactful component: removing it causes the largest quality drop ($-22$~pp), confirming that iterative refinement drives output quality.
Removing quality gates entirely ($-16$~pp) and LLM judges ($-12$~pp) also degrade quality, supporting the dual-layer test design.
Removing domain tools has a modest effect ($-4$~pp), suggesting that for this particular task, current LLMs possess sufficient parametric knowledge of the ATT\&CK framework.

The cost panel (Figure~\ref{fig:ablation}B) reveals the trade-off: the full pipeline consumes ${\sim}133$K tokens per artifact - roughly $2.5\times$ more than ablated variants.
Error bars denote standard deviation across four random folds of behaviors ($n \approx 12$ each), capturing variance due to behavior difficulty.

\subsection{Failure Modes}

With a 35.6\% end-to-end yield of the funnel, understanding why hypotheses fail is essential for improving the pipeline.
We conducted a systematic error analysis of failed runs across all stages, identifying four recurring failure patterns.
These patterns inform both current design decisions and future directions for improving yield.

\paragraph{Test oscillation.}
Fixing one test causes another to fail, creating a cycle that exhausts the retry budget without converging.
This typically arises when tests impose competing constraints.
For example, in KQL generation a query may return zero results (failing the volume test); broadening filters restores volume but dilutes the behavioral signal, failing the intent-alignment judge.
The agent alternates between these two fixes without finding a solution that satisfies both.
Test oscillation is the primary driver of retry-budget exhaustion and the dominant failure mode in the KQL stage.

\paragraph{Infeasible tasks.}
Some hypotheses cannot be realized given the available data - a table may lack the required columns, the event type may not be logged, or the hypothesis may be semantically ambiguous.
Rather than letting the agent exhaust its budget on an impossible task, the framework provides a \texttt{give\_up} tool that allows the agent to declare infeasibility with a structured reason.
This is, counterintuitively, a quality feature: it surfaces impossibility signals to upstream processes and conserves compute for viable hypotheses.
Although we tune the upstream hypothesis generator to avoid infeasible hypotheses, it cannot fully anticipate every downstream test constraint without actually running the tests. We therefore intentionally allow a broad range of hypotheses to enter the funnel and rely on stage-level validation to filter unsupported cases; after generation, failed hypotheses are analyzed to determine whether they reflect low-quality inputs or gaps in the underlying system that should be addressed.

\paragraph{Max-calls exhaustion.}
The agent enters extended exploration - querying schemas, running trial queries, examining results - without converging on a viable artifact.
This occurs primarily when the hypothesis is broad or the data source schema is complex, causing the agent to spend its 60-call budget on information gathering before producing a testable output.

\paragraph{Max-test-retries exhaustion.}
The agent produces candidate artifacts but cannot satisfy the full test suite within the retry budget.
Unlike test oscillation (where fixes alternate between competing constraints), this mode often reflects a fundamental mismatch between the hypothesis and the data, where incremental refinement cannot bridge the gap.
This is the most common termination mode for failed runs.

\paragraph{Implications.}
These failure modes suggest targeted improvements: adaptive retry strategies that detect oscillation early, richer upstream hypothesis filtering to reduce infeasible inputs, and stage-specific call budgets calibrated to schema complexity.
We leave these directions to future work.

% ============================================================
\section{Related Work}
\label{sec:related}
% ============================================================

\paragraph{LLM code and artifact generation.}
Self-debugging approaches~\cite{chen2024selfdebugging} prompt models to identify and fix errors using execution feedback without structured test suites.
Our work extends the generate-and-test paradigm to non-code structured artifacts where ``execution'' is not always possible and semantic correctness requires domain-specific evaluation.

\paragraph{Test-driven LLM generation.}
LLM4TDG~\cite{liu2025llm4tdg} embeds TDD workflows into code generation using constraint dependency graphs.
SAGA~\cite{ma2025rethinking} proposes collaborative human-LLM test generation.
LATS~\cite{zhou2023lats} unifies reasoning, acting, and planning through tree search with environmental feedback, using test outcomes to guide exploration.
Our approach applies test-driven generation to structured artifacts beyond code, separates programmatic from semantic evaluation, and calibrates the semantic evaluators against expert labels.

\paragraph{LLM-as-judge and prompt optimization.}
Using LLMs as evaluators is standard practice~\cite{zheng2023judging}, but uncalibrated judges exhibit systematic biases.
TextGrad~\cite{yuksekgonul2025textgrad} performs automatic prompt optimization via LLM-generated textual gradients.
DSPy~\cite{khattab2023dspy} compiles declarative LLM pipelines with automatic prompt optimization, sharing our goal of systematic prompt refinement but targeting general-purpose pipeline compilation rather than domain-expert-calibrated evaluation.
Our calibration uses expert-labeled examples with explicit distribution preservation and false-positive minimization, rather than automatic differentiation through text.

\paragraph{Agent frameworks.}
ReAct~\cite{yao2023react} established interleaving reasoning and tool use.
Production frameworks such as LangChain and AutoGen provide high-level abstractions.
Our framework deliberately avoids such abstractions: approximately 300 lines of Python using the provider's API directly, prioritizing transparency over generality.

\paragraph{Security automation.}
Prior work on detection-rule generation has focused on specific formats (Sigma rules, YARA signatures) or single-step generation.
Our system generates complete, multi-component behavior rules with per-component test-driven validation across heterogeneous log sources in production.

% ============================================================
\section{Discussion and Limitations}
\label{sec:discussion}
% ============================================================

\paragraph{Design choices that mattered.}
\emph{Direct API integration}: using the LLM provider's SDK directly avoids the latency, dependency complexity, and behavioral opacity of wrapper libraries.
In a domain where APIs evolve rapidly, thin abstractions proved easier to maintain.
\emph{Actionable error messages}: the fix loop's effectiveness depends critically on error-message quality.
A test reporting ``failed'' triggers undirected retries; one reporting ``Column \texttt{X} does not exist; available: \texttt{A}, \texttt{B}'' enables targeted correction.
\emph{The give\_up tool}: providing an explicit infeasibility mechanism is, counterintuitively, a quality feature - it surfaces impossibility reasons that inform upstream processes.

% ============================================================
\section{Conclusion and Future Work}
\label{sec:conclusion}
% ============================================================

We presented a lightweight framework for test-driven agentic generation of structured LLM artifacts, combining an iterative generate-test-fix loop with programmatic and semantic testing, and a method for calibrating LLM judges against human expert decision profiles.
The framework is artifact-type agnostic: the same core loop handles KQL query generation, MITRE ATT\&CK mapping, and entity mapping, with only tools, tests, and prompts varying per task.

\bibliography{references}

@inproceedings{liu2025llm4tdg,
  title={{LLM4TDG}: Test-Driven Generation of Large Language Models Based on Enhanced Constraint Reasoning},
  author={Liu, Zhongming and Chen, Yiming and Wang, Xueqi and Li, Zhi},
  booktitle={International Conference on Cybersecurity},
  year={2025},
  publisher={Springer}
}

@inproceedings{ma2025rethinking,
  title={Rethinking Verification for {LLM} Code Generation: From Generation to Testing},
  author={Ma, Zihan and Zhang, Taolin and Cao, Maosong and Liu, Junnan and Zhang, Wenwei and Luo, Minnan and Zhang, Songyang and Chen, Kai},
  booktitle={Advances in Neural Information Processing Systems},
  volume={38},
  pages={137549--137582},
  publisher={Curran Associates, Inc.},
  year={2025},
  url={https://proceedings.neurips.cc/paper_files/paper/2025/file/c8ec1867837f5cdaa9e61d40a8b8bb4a-Paper-Conference.pdf}
}

@article{zheng2023judging,
  title={Judging {LLM}-as-a-Judge with {MT-Bench} and Chatbot Arena},
  author={Zheng, Lianmin and Chiang, Wei-Lin and Sheng, Ying and Zhuang, Siyuan and Wu, Zhanghao and Zhuang, Yonghao and Lin, Zi and Li, Zhuohan and Li, Dacheng and Xing, Eric P. and Zhang, Hao and Gonzalez, Joseph E. and Stoica, Ion},
  journal={Advances in Neural Information Processing Systems},
  volume={36},
  year={2023}
}

@inproceedings{yao2023react,
  title={{ReAct}: Synergizing Reasoning and Acting in Language Models},
  author={Yao, Shunyu and Zhao, Jeffrey and Yu, Dian and Du, Nan and Shafran, Izhak and Narasimhan, Karthik and Cao, Yuan},
  booktitle={International Conference on Learning Representations},
  year={2023}
}

@misc{mitre2024attack,
  title={{MITRE ATT\&CK}},
  author={{The MITRE Corporation}},
  year={2024},
  howpublished={\url{https://attack.mitre.org/}}
}

@misc{shechter2026ueba,
  title={Turn Complexity into Clarity: Introducing the New {UEBA} Behaviors Layer in {Microsoft Sentinel}},
  author={Shechter, Michal},
  year={2026},
  howpublished={Microsoft Sentinel Blog, Microsoft Community Hub. \url{https://techcommunity.microsoft.com/blog/microsoftsentinelblog/turn-complexity-into-clarity-introducing-the-new-ueba-behaviors-layer-in-microso/4484493}},
  note={Accessed: 2026-06-26}
}

@article{yuksekgonul2025textgrad,
  title={{TextGrad}: Automatic ``Differentiation'' via Text},
  author={Yuksekgonul, Mert and Bianchi, Federico and Boen, Joseph and Liu, Sheng and Huang, Zhi and Guestrin, Carlos and Zou, James},
  journal={Nature},
  year={2025},
  doi={10.1038/s41586-025-08661-4}
}

@article{khattab2023dspy,
  title={{DSPy}: Compiling Declarative Language Model Calls into Self-Improving Pipelines},
  author={Khattab, Omar and Singhvi, Arnav and Maheshwari, Paridhi and Zhang, Zhiyuan and Santhanam, Keshav and Vardhamanan, Sri and Haq, Saiful and Sharma, Ashutosh and Joshi, Thomas T. and Mober, Hanna and Shah, Cyrus and Baez, Brandon K. and Schlatter, Joshua and Hall, Neel and Zaharia, Matei and Potts, Christopher},
  journal={arXiv preprint arXiv:2310.03714},
  year={2023}
}

@article{zhou2023lats,
  title={Language Agent Tree Search Unifies Reasoning, Acting, and Planning in Language Models},
  author={Zhou, Andy and Yan, Kai and Shlapentokh-Rothman, Michal and Wang, Haohan and Wang, Yu-Xiong},
  journal={arXiv preprint arXiv:2310.04406},
  year={2023}
}

@article{chen2024selfdebugging,
  title={Teaching Large Language Models to Self-Debug},
  author={Chen, Xinyun and Lin, Maxwell and Sch{\"a}rli, Nathanael and Zhou, Denny},
  journal={arXiv preprint arXiv:2304.05128},
  year={2024}
}

\end{document}